\documentclass[10pt,5p,twocolumn,english,preprint,times,numbers,sort&compress]{elsarticle}

\usepackage{graphicx}
\usepackage[english]{babel}
\usepackage{amsmath}
\usepackage{amssymb}
\usepackage{amsfonts}
\usepackage{longtable}
\usepackage{dcolumn}
\usepackage{bm}
\usepackage{bbm}
\usepackage{color}
\usepackage{array}
\usepackage{colortbl}

\usepackage{bbold}
\usepackage{ctable}
\usepackage{ulem}

\usepackage{mathrsfs}
\usepackage{braket}

\DeclareMathOperator{\Tr}{Tr}

\DeclareMathOperator{\e}{e}

\DeclareMathOperator{\sgn}{sgn}

\newcommand{\p}{_\text{p}}
\newcommand{\tp}{{\tau_{\mathrm{p}}}}
\newcommand{\dt}{\mathrm{d}t}
\newcommand{\F}{_\text{F}}

\begin{document}

\begin{frontmatter}

\journal{Journal of Magnetic Resonance}

\title{
Anomalous behavior of control pulses in presence of noise with singular
  autocorrelation}

\author[do]{Daniel Stanek}
\address[do]{Lehrstuhl f\"{u}r Theoretische Physik I, TU Dortmund, 
Otto-Hahn Stra\ss{}e 4, 44221 Dortmund, Germany}

\author[do]{Benedikt Fauseweh\corref{cor1}}
\ead{benedikt.fauseweh@tu-dortmund.de}

\author[kit]{Christopher Stihl}
\address[kit]{Institut f\"ur
Angewandte Materialien - IAM-AWP, Hermann-von-Helmholtz-Platz 1, 76344
Eggenstein-Leopoldshafen} 

\author[fzj]{Stefano Pasini\corref{cor1}}
\ead{s.pasini@fz-juelich.de}
\address[fzj]{J\"ulich Centre for Neutron Science, JCNS, Forschungszentrum J\"ulich GmbH, 52425 J\"ulich, Germany} 

\author[do]{G\"otz S. Uhrig\corref{cor1}}
\ead{goetz.uhrig@tu-dortmund.de}

\cortext[cor1]{Corresponding authors}

\begin{abstract}
We report on the anomalous behavior of control pulses for spins
under spin-spin relaxation and
subject to classical noise with a singular autocorrelation function. This behavior is
not detected for noise with analytic autocorrelation functions.
The effect is manifest in the different scaling behavior of
the deviation of a real pulse to the ideal, instantaneous one. While a standard
pulse displays scaling $\propto \tau_\mathrm{p}^1$, a first-order refocusing pulse
normally shows scaling $\propto \tau_\mathrm{p}^2$. But in presence of cusps
in the noise autocorrelation the scaling $\propto \tau_\mathrm{p}^{3/2}$ occurs.
Cusps in the autocorrelation are characteristic for fast
fluctuations in the noise with a spectral density of Lorentzian shape. We prove that the
anomalous exponent cannot be avoided; it represents a fundamental
limit. On the one hand, this redefines
the strategies one has to adopt to design refocusing pulses.
On the other hand, the anomalous exponent, if found in experiment,
provides important information on the noise properties.
\end{abstract}

\end{frontmatter}

\section{Introduction}

The interaction of open quantum systems with an environment, a so-called bath, is the main reason for the
decoherence of  quantum systems and the loss of well-defined phase relations. 
Especially in high precision nuclear magnetic resonance (NMR) as well as in quantum information processing (QIP)
the decoherence leads to a detrimental loss of signal strength, which is crucial in experimental measurements and in the application of quantum logic gates.
Therefore the suppression of decoherence is a long standing issue of great experimental relevance.
The two main processes of decoherence are dephasing (spin-spin relaxation, its rate defined
by the inverse time $1/T_2$) and longitudinal 
relaxation (spin-lattice relaxation, its rate defined
by the inverse time $1/T_1$). In the present paper we focus on dephasing, i.e., spin-spin relaxation.
Thereby, we implicitly assume $T_1$ to be infinite, i.e., in practice this means $T_1\gg T_2$.
But qualitatively, our conclusions pertain also to the 
case where  $T_1$ and  $T_2$ are of similar magnitude.

One step to overcome dephasing is the application of suitable control pulses to invert the time evolution 
of the spin or quantum bit. This approach dates back to the spin echo of Erwin Hahn in the fifties \cite{hahn50} 
which was quickly extended to periodically applied pulses by Carr and Purcell
\cite{carr54} and Meiboom and Gill \cite{meibo58}. Since then a large variety
of periodic pulse sequences has been proposed \cite{haebe76,mauds86,gulli90,levit05}.
In QIP, this approach is known under dynamic decoupling (DD) \cite{viola98,ban98,viola99a}.

Examples of non-periodically repeated cycles of
  ideal, instantaneous pulses are, e.g., the
recursively concatenated sequences (CDD) and the
  Uhrig dynamic decoupling (UDD) for the suppression of dephasing.  
The former require an exponentially growing
number of pulses \cite{khodj07} if longer times have to be reached,
the latter is much more efficient  because it
requires only a linearly growing number of pulses 
\cite{uhrig07,uhrig07err,uhrig08,uhrig08err,yang08b}. UDD pays for
power spectra with hard cutoff \cite{cywin08,uhrig08,uhrig08err,pasin10a},
while for soft cutoffs the Carr-Purcell-Meiboom-Gill sequence works
best \cite{ajoy11}.
Experimental verification of these theoretical proposals
 are available \cite{bierc09a,du09,alvar10c}.
 Furthermore, UDD and other non-periodic sequences have been used
 to enhance the contrast in magnetic resonance imaging \cite{jenis09, stokes12}.

Experimentally pulses always have a finite duration $\tp$
and they are susceptible to the perturbing influence of the bath.
This influence can be reduced by using composite pulses 
\cite{tycko83,levit86,cummi00,cummi03,alway07},
by shaping the pulse \cite{pasin08a,pasin08b,pasin09a,fause12},
 or by applying dynamically corrected gates
 \cite{khodj09a,khodj10,bierc13}. Of course, shaping of pulses has also been 
a long-standing issue in NMR. The main aim
was to generate \textit{robust} pulses which are only weakly sensitive to pulse imperfections.
We can only mention a small part of the abundant literature on this issue
\cite{tycko83,warren84,warren85,levit86,lerner93,lerner95,cummi00,cummi03,skinn03,kobza04,sengu05,motto06,alway07,pryad08a,pryad08b,souza12}, for a book see Ref.\ \cite{levit05}.

In most experiments of NMR the dominating fluctuations
destroying coherence are of \textit{classical}
nature, see, e.g., Ref.\ \cite{bierc11b}. Generically the decohering fluctuations 
are induced by a large number of microscopic and even macroscopic degrees of freedom.
At least a part of these degrees of freedom are thermalized at rather high temperatures relative to their 
generic energy scales. For instance, the nuclear spins can be considered
to be in a completely disordered state corresponding to infinite temperature. 
Thus the resulting fluctuations are thermal fluctuations and their quantum character
plays only a minor role \cite{stanek2013,hackm14b} so that it can be neglected.

One may think that a strongly disordered state of the decohering bath poses a problem
to the preservation of coherence in the systems under study. But the opposite is true:
The lack of quantumness of the fluctuations allows us to consider classical fluctuations.
This represents a simplification because the classical variables
always commute. Hence, the necessary pulse shapes are simpler and the required amplitudes
are lower than in the full quantum case. 
Moreover, the simulation of the decoherence is computationally much simpler.

In this article we present simulations of pulses, which are optimized with respect to their robustness
against dephasing from classical baths or from quantum baths.
We simulate a single spin coupled to a classical dephasing bath causing 
spin-spin relaxation and investigate how the autocorrelation $g(t)$ of the bath affects the performance of the pulses.
Our primary finding is that the possibility to optimize the pulse shape depends qualitatively on the autocorrelation function of the noise.
For analytic autocorrelation functions, e.g., a Gaussian function $\log g(t) \propto -t^2$, the performance of the pulses depends on its shape. 

Theoretically, this is expressed by the scaling property of the deviation 
of the real pulse from the ideal one with respect
to the duration $\tp$ of the pulse. A standard
unshaped pulse displays scaling $\propto \tau_\mathrm{p}^1$ and an $m^\text{th}$ order refocusing pulse, appropriately
shaped, displays the  scaling $\propto \tau_\mathrm{p}^{m+1}$. Solutions for $m=1$ and $m=2$ are known \cite{cummi00,cummi03,sengu05,pryad08a,pryad08b,pasin08a,pasin09a,khodj10,fause12}.
If, however, the autocorrelation of the noise displays a cusp at
$t=0$, e.g., $\log g(t) \propto -\left|t\right|$, there is \textit{no}
pulse shape which makes scaling $\propto \tau_\mathrm{p}^{2}$ possible. The best
possible results is an anomalous scaling $\propto \tau_\mathrm{p}^{3/2}$.
This is the key result of our present work. We will prove it mathematically
and illustrate it in simulations.

We stress that our finding is not an academic secondary aspect.
The exponential decaying autocorrelation is
generic for many processes: For instance, it is characteristic for the
omnipresent Ornstein-Uhlenbeck process indicating the slow decay of a quantum state
into a continuum of states reaching up to  high energies. 
The best known popular example is radioactive
decay. In the context of general spin decoherence we refer the reader to Refs.\ \cite{klaud62,lange10}.
After a Fourier transform, an Ornstein-Uhlenbeck noise displays
a Lorentzian spectral density. Generally, the cusp of $g(t)$ at $t=0$ yields
a slow asymptotic decay $\propto 1/\omega^2$ for large frequencies.
In NMR, noise spectral densities of Lorentzian form are also common \cite{abrag78},
leading to homogeneous broadening of the NMR lines.
Microscopically, such noise is generated by the fluctuations of dipolar interactions 
due to random molecular rotations and translations \cite{abrag78}. 
These rotations are generally very fast. Often, they also imply $T_1\approx T_2$.
But in large molecules, e.g., bio-molecules, the time
scales can be such that $T_1 \gg T_2$. Moreover, we will argue below that the
fundamental limits to pulse refocusing are valid also if $T_1$ and
$T_2$ are of similar magnitude.

The paper is organized as follows: In Sect.\ \ref{sec:systems} the model 
under study and the ansatz for the shaped pulses is presented.  
In Sect.\ \ref{sec:FN} a measure for the difference between the 
real and the ideal pulse is introduced, the Frobenius norm,
and the requirements for first- and second-order shaped pulses are derived.
In Sect.\ \ref{sec:simulation} we report the results of our simulations for an analytical  and 
a cusp-like autocorrelation function.
In Sect.\ \ref{sec:cusp} we explain why the unexpected $3/2$ scaling occurs and prove the no-go theorem for cusp-like autocorrelation functions.
Finally, we conclude in Sect.\ \ref{sec:conclusions}.

\section{Classical noise systems}
\label{sec:systems}

We restrict ourselves to pulses in presence of pure dephasing (spin-spin relaxation). 
This approximation can be used wherever $T_1\gg T_2$ holds, for instance
in large magnetic fields.  The corresponding  Hamiltonian reads in the rotating frame
\begin{subequations}
\begin{equation}
\label{eq:hamiltonian_tot}
H_\text{tot}= H(t) + H_\text{p} (t)
\end{equation}
with the control Hamiltonian
\begin{equation}
\label{eq:hamiltonian_p}
H_\text{p}(t)= v(t) \sigma_x
\end{equation}
 and
\begin{equation}
\label{eq:hamiltonian_sys}
H(t)= \eta(t) \sigma_z
\end{equation} 
\end{subequations}
the Hamiltonian for the classical noise. Here we use the
term ``classical'' to indicate that $\eta (t)$ is a
random variable of real numbers obeying a Gaussian distribution.
There is no back-action from the spin to the noise and since it
couples only to the $\sigma_z$ operator this coupling commutes with possible
drift terms in the Hamiltonian.
The Gaussian distribution
is defined by its one-point and two-point correlations, i.e.,
by $\eta_0(t)=\overline{\eta(t)}$ and by $g(t_2,t_1):=\overline{\eta(t_2)\eta(t_1)}$.
In addition, it is highly plausible to assume that the bath is translationally invariant
in time, i.e., any shift $t\to t+\Delta t$ does not alter the noise.
This implies that $\eta_0$ is constant in time and that the autocorrelation
$g(t_2,t_1)$ depends only on the time difference $g(t_2-t_1):=g(t_2,t_1)$.
The value of $\eta_0$ can be seen as an additional magnetic field.

If the random field changes only slowly in time the autocorrelation $g(t)$ will
decay slowly. If it changes quickly $g(t)$ will decay quickly.
If $\eta(t)$ can change rather abruptly, this corresponds to the 
influence of noise components at high frequencies. Consequently,
the decay in $g(t)$ displays a singularity at $t=0$.

In the control Hamiltonian, $v(t)$ stands for the
amplitude of the pulse used to rotate the spin around the $x$
axis. The angle of rotation is given by the time integral of the pulse
amplitude  
\begin{equation}
\psi(t) := 2 \int_0^{t} \text d t'\ v(t').
\end{equation}
The full angle of the pulse is given by $\psi(\tp)$.
For clarity, we consider here
spin-flip pulses only with $\psi(\tp)=\pi$.

The finite duration of the pulse and the non-commutativity of 
$H(t)$ and $H\p(t)$ make the evolution of the combined system non-trivial. 
We stress, however, that it remains a unitary evolution given by
the unitary evolution operator $U_\text{{tot}}(\tp)=
{\cal T}\exp\{-i \int_0^\tp H_\text{tot}(t)\text d t\}$. It cannot 
be written as the product of the time evolution
of the bath alone and the pulse operator $P_{\tp}$.
This form would only hold for an infinitely short pulse
because the coupling to the bath would not matter. Formally, one has
\begin{equation}
\label{eq:aim1}
U_\text{tot}(\tp) = P_{\tp} + {\cal O}(\tp).
\end{equation} 

A more sophisticated solution consists in shaping the pulse such that
the above approximation holds in the form
\begin{equation}
\label{eq:aim2}
U_\text{tot} (\tp) = P_\tp + {\cal O}(\tau_\mathrm{p}^{m+1})
\end{equation} 
with integer $m\ge 0$. We call such a pulse an $m$th-order pulse. 
Thus, a standard unshaped pulse is a zeroth-order pulse.
In this paper, we work with first- and second-order pulses.

The ansatz \eqref{eq:aim2} can be realized for arbitrary $m$ in principle \cite{khodj10} and concrete
proposals have recently been made for second order pulses ($m=2$)
for quantum baths \cite{pasin09a,fause12}. Furthermore, it could be shown
that pulses with this property can be used in pulse sequences if they are not
applied at the same instants as instantaneous pulses would be applied, but in adapted sequences
\cite{uhrig10a,pasin11a}. These observations corroborate that the shaping of pulses
is an important ingredient for high-fidelity coherent control.

It is not the scope of the present paper to describe how one can find such pulses,
see for instance Refs.\ \cite{tycko83,warren84,cummi00,cummi03,pryad08a,pryad08b,pasin09a,pasin11a,fause12}.
Our point here is that the performance of the refocusing pulses 
depends on the nature of the bath, i.e., on the qualitative form of its
autocorrelation function.  For illustration, we simulate autocorrelation functions with either a Gaussian
or an exponential decay. 

\section{Frobenius norm as measure of pulse quality}
\label{sec:FN}

The Frobenius norm as such is a way to define a modulus for matrices \cite{bhati97}.
It is unbiased in the sense that the Frobenius norm does not depend on the basis in which the
matrix is given as long as it is an orthonormal one. We use 
the Frobenius norm as a measure for the quality of the pulse
by applying it to the difference between the density matrix subject to
the ideal pulse and the one subject to the real pulse. 

We will exploit below that the random dynamics induced by the
coupling to classical noise preserves the unitarity of the time evolution
so that the Frobenius norm reflects the same behavior as the
deviation of a fidelity from unity. Moreover, we will illustrate that
the Frobenius norm reflects the behavior of various polarized states
so that its behavior is representative for the time evolution of 
generic initial states.

\subsection{General properties}
\label{subsec:genprop}

To be specific, we consider the initial density matrices for the completely polarized
spin states
\begin{equation}
\label{eq:rho_0}
\rho_0^\alpha= \frac{1+\sigma^\alpha}{2} 
\end{equation}
where $\alpha=x,y$ or $z$ defines the direction of the polarization
and $\sigma^\alpha$ are the Pauli matrices. Next, we let the spin
evolve under the evolution operator $U_\text{tot}(\tp)$ 
so that the final density matrices are
$\rho^\alpha_1:=U_\text{tot}(\tp)\rho^\alpha_0 U_\text{tot}^\dagger (\tp)$.
They are compared to the ideal outcomes
$\rho^\alpha_\text{id}:=P_{\tp} \rho^\alpha_0 P_{\tp}^\dagger$
which have evolved under  the perfect pulse $P_{\tp}$.
Thus, we define the Frobenius norm $\Delta\F$ by
\begin{equation}
\label{eq:Frobenius_norm}
\Delta_\text{F}^2:=\frac{1}{3}\sum_{\alpha=x,y,z}\text{{Tr}}\left(\rho^\alpha_\text{id}-\rho^\alpha_1\right)^2.
\end{equation}
The sum over the three orthogonal directions is introduced to
keep the definition independent from the polarization of the initial state.

Using general properties of density matrices we obtain
\begin{equation}
\label{eq:Frobenius_norm_simp}
\Delta_\text{F}^2 = 2\left[1-\frac{1}{3}\sum_{\alpha=x,y,z}\Tr
\left(\rho^\alpha_\text{id}\rho^\alpha_1\right)\right].
\end{equation} 
Note that the sum in this equation is the fidelity of the real pulse.
Its difference to unity yields  the Frobenius norm up to a factor of 2.
For later use, we define the partial Frobenius norm relevant for the individual
directions
\begin{equation}
\label{eq:Frob_partial}
\left(\Delta_\text{F}^{(\alpha)}\right)^2 := 2\left[1-\Tr\left(\rho^\alpha_\text{id}\rho^\alpha_1\right)\right]
\end{equation} 
so that $\Delta_\text{F}^2=1/3\sum_{\alpha=x,y,z}(\Delta_\text{F}^{(\alpha)})^2$.

Further simplification can be achieved by the general ansatz
\begin{equation}
\label{eq:genAnsatz}
U_\text{tot}(\tp)=P_{\tp}U_\text{c}(\tp)
\end{equation}
where all corrections of the real evolution $U_\text{tot}(\tp)$
relative to the ideal one $P_{\tp}$ are put into the correcting factor $U_\text{c}(\tp)$.
It is given by 
\begin{subequations}
\begin{eqnarray}
\label{eq:Uc}
U_\text{c}(\tp) &=& {\cal T}\left[\exp\left(-i \int_0^{\tp} \widetilde H(t)\text d t\right)\right]
\\
\label{eq:Htilde}
\widetilde H(t) &:=& P_{t}^\dagger H(t)P_{t}
\end{eqnarray}
\end{subequations}
where $P_t$ is the unitary time evolution induced by the control Hamiltonian
$H\p(t)$ alone.
As was to be expected,  the correcting factor depends solely on the coupling between the spin
and the bath. By exploiting the unitarity of $P_{\tau_\text{p}}$ and the
properties of the trace, the ideal rotations in $\Delta_\text F$
cancel and the Frobenius norm reduces to 
\begin{equation}
\label{eq:FN_reduced}
\Delta\F^2= 2-\frac{2}{3} \sum_{\alpha=x,y,z} \Tr 
\left( \rho_0^\alpha U_\text c(\tp)\rho_0^\alpha U^\dagger_\text c(\tp)
\right).
\end{equation}
Inserting the initial states and expanding yields
\begin{equation}
\label{eq:FN_reduced2}
\Delta\F^2=1-\frac{1}{6} \sum_{\alpha=x,y,z} \Tr
\left( \sigma^\alpha U_\text c(\tp)\sigma^\alpha U^\dagger_\text c(\tp)
\right).
\end{equation}

The correcting factor $U_c$ is a unitary in the two-dimensional Hilbert space of a
spin $S=1/2$. Thus it can always be written as 
\begin{equation}
\label{eq:Magnus_Uc}
U_\text{c}(\tp) = \exp(-\text{i} \vec\mu\cdot \vec\sigma).
\end{equation}
Using Euler's formula this can be expanded yielding
\begin{equation}
\label{eq:euler}
 U_\text{c}(\tp) = \cos \left| \vec{\mu} \right| - \text{i} \frac{\vec{\mu} \cdot \vec{\sigma}}{\left| \vec \mu \right|}\sin \left| \vec{\mu} \right|.
\end{equation}
Inserting \eqref{eq:euler} into \eqref{eq:FN_reduced2} we obtain 
\begin{equation}
\label{eq:FN_vs_mu}
\Delta\F^2=\frac{4}{3}\left[1-\cos^2|\vec\mu |\right].
\end{equation}
This simple relation shows that the deviation of the real pulse from the
ideal one only depends on the deviation of the correcting unitary $U_c$ from
the identity. No direction in spin space is singled out, i.e., the expression is
fully rotationally symmetric in spin space. Since we are mostly interested in the
case of small deviations it is sufficient to consider
\begin{equation}
\label{eq:FN_vs_mu2}
\Delta\F^2=\frac{4}{3} \vec\mu ^2 + {\cal O}(|\vec\mu |^4).
\end{equation}

\subsection{Average Hamiltonian theory}
\label{subsec:AHT}

The vector $\vec\mu$ can be expressed  as the sum 
$\vec\mu=\sum_{k=1}^\infty \vec\mu^{(k)}$
resulting of the application of a Magnus expansion
\cite{blane09}, see also Ref.\ \cite{warren84},
to the time-dependent Hamiltonian $\widetilde H(t)$ in
\eqref{eq:Htilde}. Each contribution $\vec\mu^{(k)}$
results from a $k$-dimensional integral over $(k-1)$-fold nested commutators of $\widetilde H(t)$
at different times. So $\vec\mu^{(k)} = {\cal O}(\tau_\mathrm{p}^k)$ holds.
The leading  terms  read
\begin{subequations}
\label{eq:Magnuns_mu1}
\begin{align}
\label{eq:Magnus_mu1y}
&\mu_y^{(1)}=\int_0^{\tau_\text p}\text dt\, \eta(t) \sin\psi(t)\\
\label{eq:Magnus_mu1z}
&\mu_z^{(1)}=\int_0^{\tau_\text p}\text dt\, \eta(t) \cos\psi(t).
\end{align} 
The non-vanishing second order term is
\begin{align}
\label{eq:Magnus_mu2x}
\mu_x^{(2)}=\int_0^{\tp}\!\text{d}t_1 \int_0^{t_1}\! \text
dt_2 \, \eta(t_1)\eta(t_2) \sin\left[\psi(t_1)-\psi(t_2) \right].
\end{align}
\end{subequations}

At this stage, one may start expanding $\eta(t)$ in a Taylor series
and trying to make each resulting term vanish. But for a random noise
variable this does not make sense. The solution suggesting itself is
to discuss the average of the square $\Delta\F^2$ of the Frobenius norm over
random time series $\eta(t)$. This is what we will do numerically 
in the next section in the simulations. In principle, one can also
average $\Delta\F$ instead of the square. We found that this does not 
make a noticeable difference.

Analytically, we consider here only the leading, first order which
becomes
\begin{subequations}
\label{eq:Magnus_mu1}
\begin{align}
\label{eq:Magnus_mu1y_aver}
& \overline{\mu_y^{(1)}} =\eta_0 \int_0^{\tau_\text p}\text dt \sin\psi(t)\\
\label{eq:Magnus_mu1z_aver}
& \overline{\mu_z^{(1)}} =\eta_0 \int_0^{\tau_\text p}\text dt  \cos\psi(t).
\end{align} 
\end{subequations}
Thus, one requires a first-order pulse to fulfill
\begin{subequations}
\label{eq:first_cond}
\begin{align}
\label{eq:Magnus_mu1y_cond}
& 0 = \int_0^{\tau_\text p}\text dt \sin\psi(t)\\
\label{eq:Magnus_mu1z_cond}
& 0 = \int_0^{\tau_\text p}\text dt  \cos\psi(t).
\end{align} 
\end{subequations}
Note that these conditions are necessary for a first-order
pulse even if $\eta_0$ vanishes because in the 
average of $\Delta\F^2$ the terms $\overline{(\mu_y^{(1)})^2}$ and $\overline{(\mu_z^{(1)})^2}$ 
occur as well. Their leading order in $\tp$ is proportional to the variance of $\eta$ given by $g(0)>0$
and to the square of the integrals in \eqref{eq:first_cond}, see Sect.\ \ref{sec:cusp}.

\section{Numerical simulations}
\label{sec:simulation}

We  numerically simulated the system of Eq.\ (\ref{eq:hamiltonian_tot})
with either a Gaussian or an exponential autocorrelation function and 
for refocusing pulses of different order: The zeroth-order pulse (i.e.\ a rectangular
pulse with constant amplitude), the CORPSE and SCORPSE \cite{cummi00,cummi03} first-order pulses
and three different second-order pulses, CLASS2ND optimized for a classical bath \cite{stihl12b}
and the pulses SYM2ND and ASYM2ND originally optimized for quantum
baths \cite{pasin09a}. All these pulses are
piecewise constant $\pi$-pulses; their details are given in \ref{app:A}.
The discretization of time  in the simulation 
is adapted to represent the switching instants, i.e., the instants
at which the amplitude jumps, accurately. For simplicity, we assume
that the noise does not have any frequency offset, i.e., the simulations
are based on $\eta_0=0$.

The core part of our simulation program is the sampling of white Gaussian
noise that obeys a pre-defined autocorrelation function. For sufficient
statistics we need to average over a large number of fluctuations;
this requires a random number generator with a very large period. We
used the Mersenne Twister generator to sample a large number of discretized
 fluctuations $\bm{\eta}=(\eta_1,\ldots,\eta_N)^\top$ where $N$ is the number of time steps $t_i$.
To this end, we first generate a sequence $\bm{r}=(r_1,\ldots,r_N)$ of uncorrelated
 random numbers obeying Gaussian statistics with vanishing mean value 
and individual standard deviations
 $(\sigma_1,\ldots,\sigma_N)^\top=(\mathrm{diag}\ \mathbf{D})$. 
The diagonal matrix $\mathbf{D}$ is given by the eigen decomposition
 $\mathbf{G}=\mathbf{O}\mathbf{D}\mathbf{O}^\top$ of the covariance matrix
$\left(\mathbf{G}\right)_{ij}=g(t_i-t_j)$ of the random fluctuations
 $\eta(t)$. Finally, the vector $\bm{r}$ is transformed to the non-diagonal
 basis $\bm{\eta}=\mathbf{O}\bm{r}$ to obtain the correlated Gaussian noise
 $\bm{\eta}$. 

Since we are dealing with a two-level system, the numerical integration is
 carried out easily for arbitrary pulses. A reduction of the integration
 error for time-dependent Hamiltonians is achieved by commutator-free 
exponential time propagators (CFETs) as introduced by Alvermann
 \textit{et al.}~\citep{alver11,alver12}. 
The additional numerical effort is negligible because an exact 
analytical representation of the time-evolution operator of 
a two-level system is available. Hence, no additional diagonalizations are 
induced by the CFETs.

We render the results of our simulations as
functions of $1/v$, where $v$ denotes the maximum amplitude of the
pulse. This choice respects the scenario relevant for experiments
because generically the constraint is on the maximum realizable
amplitude, not on the duration $\tp$ of the pulse. Thus, a 
simple pulse, though of low order, may be experimentally advantageous 
because it is shorter. This feature is taken into account by the plots 
against $1/v$. We point out that for a given constant amplitude different
$\pi$-pulses will be of different durations $\tau_\text p$.
But for the same pulse shape $\tp \propto 1/v$ holds so that
the same scaling is observed.

\subsection{Noise with Gaussian autocorrelation}
\label{subsec:gaussian-g}

\begin{figure}[tb]
 \centering
  \includegraphics[width=0.7\columnwidth,angle=-90]{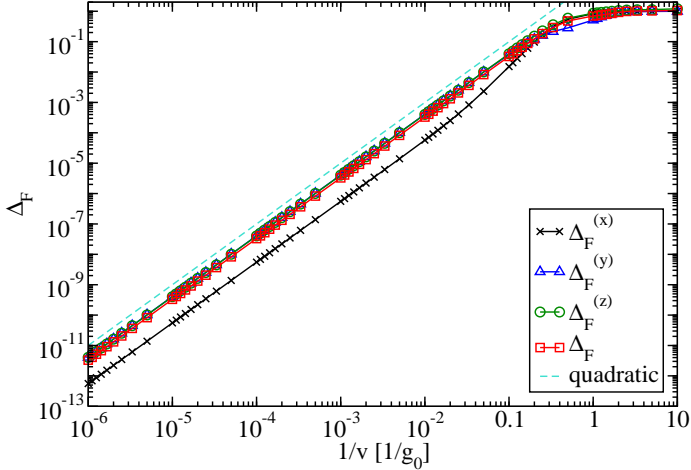}
 \caption{Frobenius norm $\Delta\F$ for a SCORPSE pulse and the individual contributions 
$\Delta\F^{(\alpha)}$ defined in Eq.\ \eqref{eq:Frob_partial} 
from  $\rho^\alpha, \alpha\in\{x,y,z\}$ versus the inverse pulse amplitude $1/v$. 
The autocorrelation function is given in Eq.\ \eqref{eq:Gaussian_g} with $\gamma=0.1\,g_0$.
}
\label{fig:Frob_SCORPSE}
\end{figure}

We consider a bath with the Gaussian autocorrelation function 
\begin{equation}
\label{eq:Gaussian_g}
g(t)=g_0^2\text e^{-\gamma^2 t^2},
\end{equation} 
where $g(0)=g_0^2$ determines the energy scale of the system while
$\gamma$ sets the decay rate of the correlation. 

In Fig.\ \ref{fig:Frob_SCORPSE}, we illustrate how the total Frobenius norm
is composed of its partial contributions. The figure
exemplarily show the results
for the partial contributions to the Frobenius norm~\eqref{eq:Frob_partial} for the SCORPSE pulse.
The central message of Fig.\ \ref{fig:Frob_SCORPSE} is that
 all three contributions display the 
\textit{same} scaling behavior $\propto \tau_\mathrm{p}^2\propto 1/v^2$.
Thus the total Frobenius norm follows the same scaling and there is no
need to consider the partial contributions separately.
Otherwise, one should discuss the full map of the real pulse
as in quantum tomography.

\begin{figure}[tb]
 \centering
  \includegraphics[width=0.7\columnwidth,angle=-90]{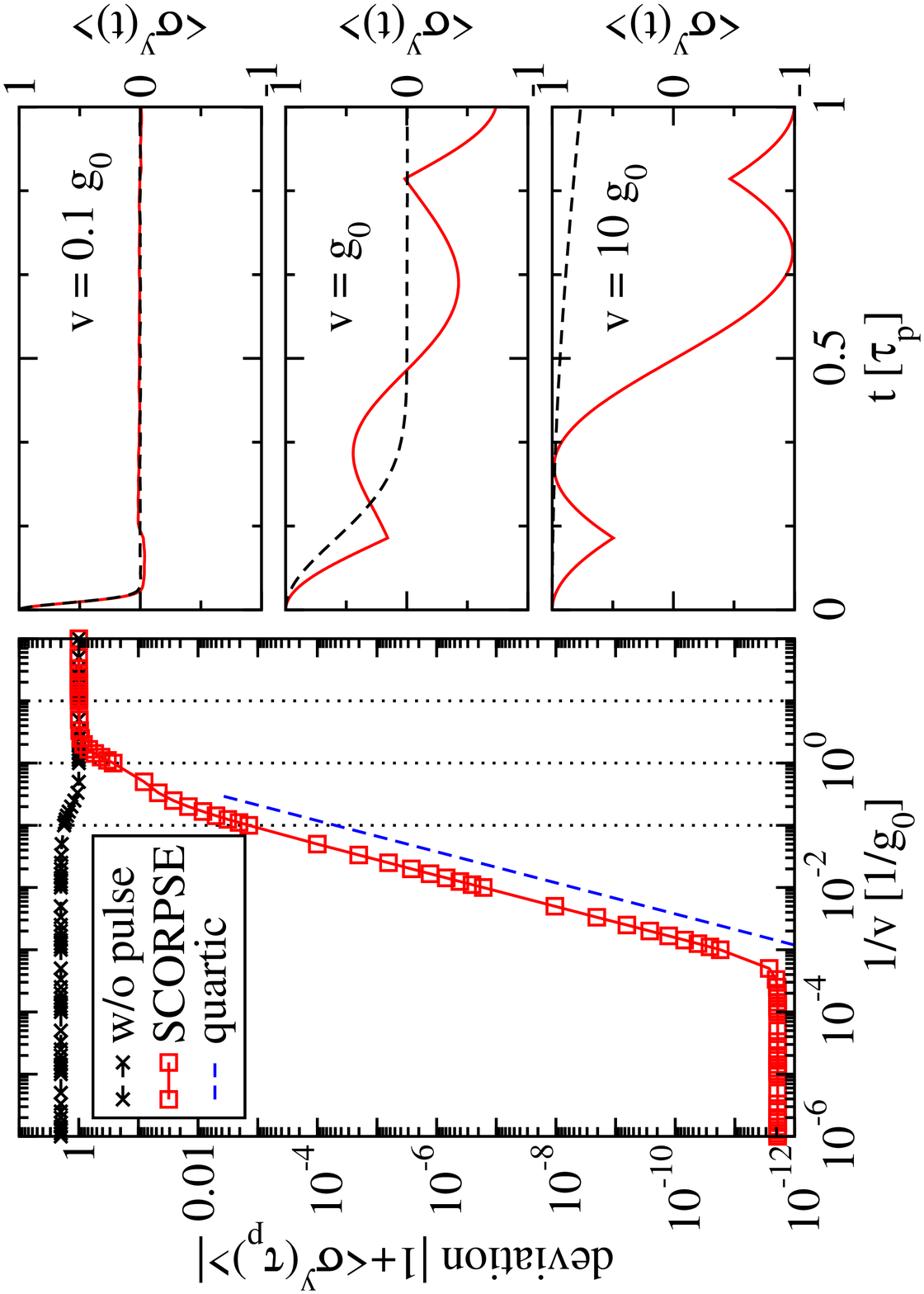}
 \caption{Deviation of the polarization $\braket{\sigma^y(\tp)}$ from its ideal value $\braket{\sigma^y(\tp)}=-1$ after the application of the SCORPSE pulse as a function of the inverse pulse amplitude $1/v$ (left panel).
In the right panel, the polarization $\braket{\sigma^y(t)}$ in the course of the SCORPSE pulse is plotted  for three different values of the pulse amplitude $v$.
The noise follows the autocorrelation \eqref{eq:Gaussian_g} with $\gamma=0.1\,g_0$.
}
\label{fig:Mag_SCORPSE}
\end{figure}

Fig.\ \ref{fig:Mag_SCORPSE} illustrates the detailed analysis of a particular polarization.
The right panels show the temporal evolution of the $y$-polarization during the pulse.
Due to the spin flip around $\sigma^x$ it is eventually negated. How well this is achieved
depends on the duration $\tp$ of the pulse or, putting the same in a different way, 
on the inverse amplitude $1/v$. For larger amplitude the wanted value $\braket{\sigma^y(\tp)}=-1$
is achieved better than for lower values. This fact is summarized in the left panel where
also the quartic scaling is shown. This quartic power law is the direct consequence
of the scaling $\Delta\F\propto \tau_\mathrm{p}^2$ because for a $\pi$-pulse a small error $\Delta\psi$ in the
spin flip angle $\psi(\tp)$ implies a deviation $\propto (\Delta\psi)^2$. Thus,
the deviation $|1+\braket{\sigma^y(\tp)}|$ scales like $\Delta\F^2\propto \tau_\mathrm{p}^4 \propto 1/v^4$.
The key point is again that the relevant scaling feature is represented by the total
Frobenius norm $\Delta\F$.

\begin{figure}
 \centering
  \includegraphics[width=0.7\columnwidth,angle=-90]{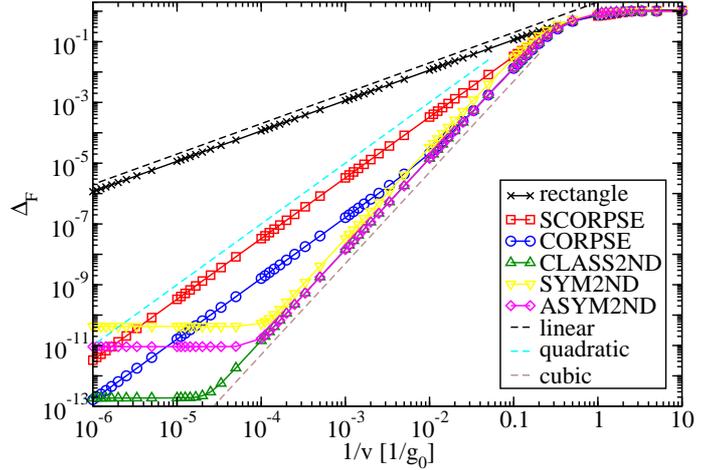}
 \caption{Frobenius norm $\Delta\F$ versus the inverse pulse amplitude $1/v$ for various pulses, see legend.
 For comparison, various power laws are depicted as dashed lines. 
The autocorrelation is Gaussian, see Eq.\ \eqref{eq:Gaussian_g}, with $\gamma=0.1\,g_0$. 
    The simulation was carried out for $500,000$ random configurations $\eta(t)$.
}
\label{fig:Frob_gaussian}
\end{figure}

On the basis of the above considerations we are now in the position to compare the scaling of $\Delta\F$
of different pulses, knowing that this measure represents the performance of the pulses.
 We do so in Fig.\ \ref{fig:Frob_gaussian}. The pulses are given in detail in \ref{app:A}.
 It is obvious that they obey significantly different scalings. As expected, the standard
 rectangular pulse displays linear scaling. The first-order pulses SCORPSE and CORPSE indeed
 display the quadratic scaling. It is remarkable, that the CORPSE pulse does even better
 in a certain transition region of larger values $1/v$. There it behaves like a second-order pulse
 with cubic scaling down till $1/v\approx 10^{-2}/g_0$, where the exponent switches to quadratic behavior.
 The second-order pulses by design (CLASS2ND, SYM2ND, ASYM2ND) display the proper cubic scaling.
 They behave very similarly so that from this simulation none of them is to be preferred over the others.
 
 At very low values of $1/v$ and thus of $\Delta\F \approx 10^{-12}$ plateaus appear for the more complicated pulses.
 This is a direct effect of numerical inaccuracies in the simulations or in the representation of the pulses.
 For lower accuracies, the plateaus would appear at larger values of $\Delta\F$. For theoretical clarity,
 we used the high-precision data for the pulses given in \ref{app:A}. This allows us to show the asymptotic power laws
 very clearly. Of course, in experiment a lower accuracy in the realization of the pulses will
 lead to plateau-like behavior for larger $\Delta\F$. 
 We observe linear scaling between the pulse accuracy and the
 plateau of $\Delta\F$, this means, that a $10^{-3}$ accuracy in the pulse realization will roughly
 imply that the power law in the pulse performance is cut off at $\Delta\F\approx 10^{-3}$ and so on.
 
The key message of  Fig.\ \ref{fig:Frob_gaussian} is that
the simulations confirm the expected analytical behavior for which
the pulses were designed. An $m$-th order pulse indeed shows scaling $\propto \tau_\mathrm{p}^{m+1} \propto 1/v^{m+1}$.
This is established exemplarily for $m=0, 1, 2$.

\subsection{Noise with exponential autocorrelation}
\label{subsec:exponential-g}

\begin{figure}[tb]
 \centering
  \includegraphics[width=0.7\columnwidth,angle=-90]{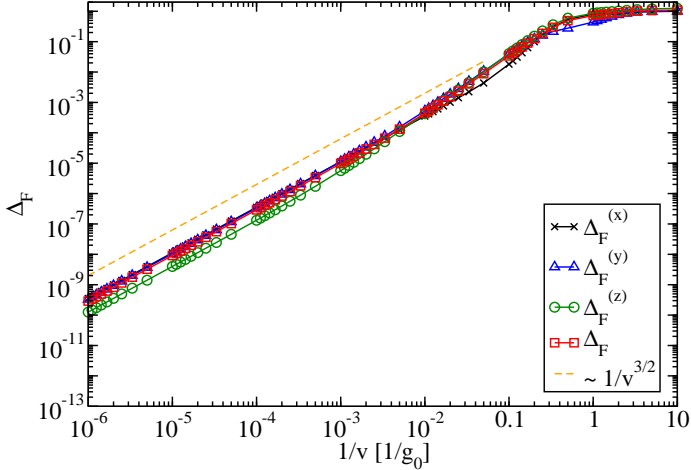}
 \caption{Frobenius norm $\Delta\F$ for a SCORPSE pulse and the individual contributions 
$\Delta\F^{(\alpha)}$ defined in Eq.\ \eqref{eq:Frob_partial} 
from  $\rho^\alpha, \alpha\in\{x,y,z\}$ versus the inverse pulse amplitude $1/v$. 
The autocorrelation function is given in Eq.\ \eqref{eq:exponential_g} with $\gamma=0.01\,g_0$.
}
\label{fig:Frob_exp_SCORPSE}
\end{figure}

We study the exponentially decaying autocorrelation function 
\begin{equation}
\label{eq:exponential_g}
g(t)=g_0^2\text e^{-\gamma |t|}.
\end{equation} 
The main difference to the Gaussian autocorrelation simulated in the  previous subsection
is the cusp at $t=0$.

\begin{figure}[tb]
 \centering
  \includegraphics[width=0.7\columnwidth,angle=-90]{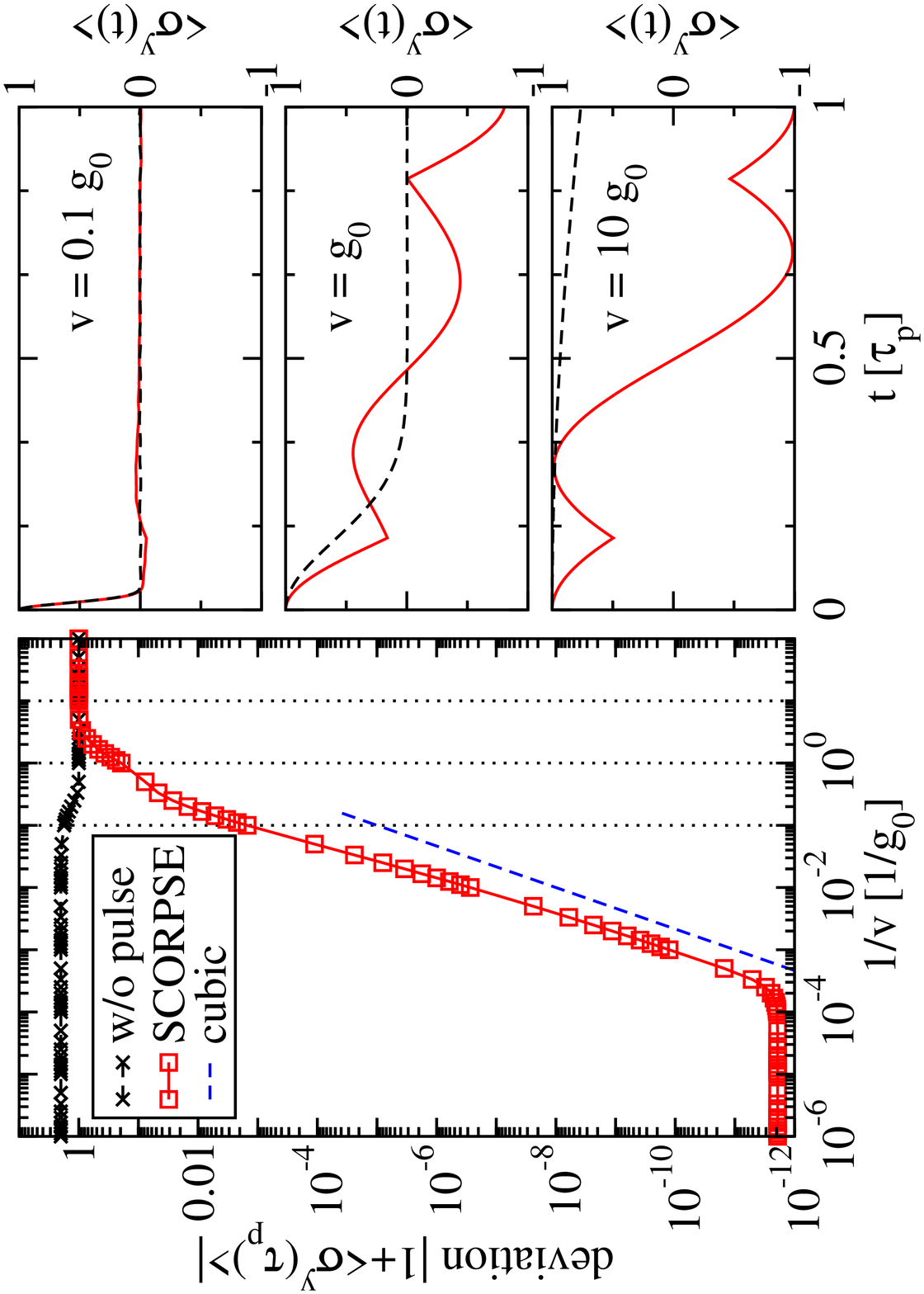}
 \caption{Deviation of the polarization $\braket{\sigma^y(\tp)}$ from its ideal value $\braket{\sigma^y(\tp)}=-1$ after the application of the SCORPSE pulse as a function of the inverse pulse amplitude $1/v$ (left panel).
In the right panel, the polarization $\braket{\sigma^y(t)}$ in the course of the SCORPSE pulse is plotted  for three different values of the pulse amplitude $v$. 
The noise follows the autocorrelation \eqref{eq:exponential_g} with $\gamma=0.01\,g_0$.
}
\label{fig:Mag_exp_SCORPSE}
\end{figure}

First, we show as before that the partial contributions along the three directions
behave the same in the sense that they display the same scaling, see Fig.\ 
\ref{fig:Frob_exp_SCORPSE}. The precise form of the scaling is indeed anomalous displaying 
the proportionality $\propto \tau_\mathrm{p}^{3/2} \propto 1/v^{3/2}$. We will come back to this point below.
Similarly, we confirm that the Frobenius norm is representative for the behavior of the polarization,
see Fig.\ \ref{fig:Mag_exp_SCORPSE}. The scaling observed in the left panel is consistent
with the scaling observed in Fig.\ \ref{fig:Frob_exp_SCORPSE}, but it differs from what we
found for the analytic autocorrelation in the previous subsection.

\begin{figure}
 \centering
  \includegraphics[width=0.7\columnwidth,angle=-90]{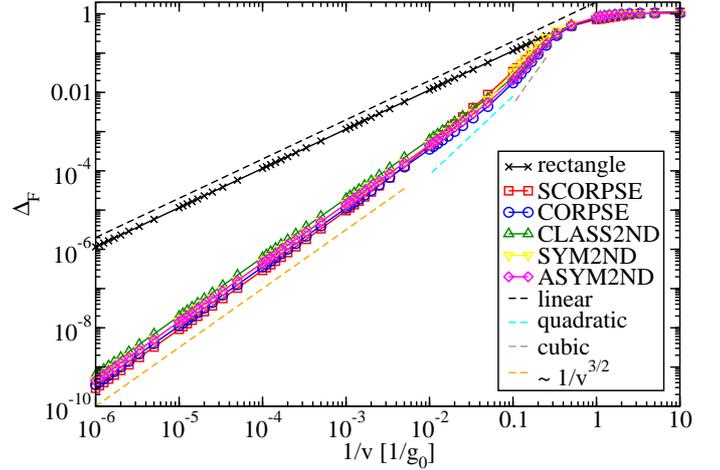}
 \caption{Frobenius norm $\Delta\F$ versus the inverse pulse amplitude $1/v$ for various pulses, see legend.
 For comparison, various power laws are depicted as dashed lines. 
The autocorrelation is an exponential, see Eq.\ \eqref{eq:exponential_g}, with $\gamma=0.01\,g_0$. 
  The simulation was carried out for $500,000$ random configurations $\eta(t)$.}
\label{fig:Frob_exp}
\end{figure}

Fig.~\ref{fig:Frob_exp} summarizes the comparison of the simulations of $\Delta\F$ of various pulses.
The unshaped rectangular pulse behaves linearly as expected. 
All pulses of first and second order reveal a rather surprising behavior expressed by the scaling
$\Delta\F\propto 1/v^{3/2}$ involving the  unexpected half-integer  value 3/2 for the scaling exponent.
Power-laws with exponents larger than 3/2 can be discerned for 
very limited regions of larger values of $1/v$.
For faster decaying autocorrelations, i.e. $\gamma > 0.01\,g_0$, these regions vanish completely and 
$\Delta\F\propto 1/v^{3/2}$ is the only identifiable power law. Thus, it is detectable also for an
accuracy of, e.g., $10^{-3}$.

To our knowledge, the occurrence of half-integer exponents in the scaling of pulse performance is a
unique feature which has not been mentioned in the literature so far. 
Similarly, we stress that it is remarkable that the first- and the second-order
pulses behave almost alike in striking contrast to what can be observed for
the Gaussian autocorrelation in Fig.\ \ref{fig:Frob_gaussian}.
It bears an important implication for experiment because it sets strict limits
to what can be achieved by pulse design.

Of course, it is important to understand the origin of this unexpected and
anomalous scaling. It is plausible to attribute it to the  singularity
in the autocorrelation. This view is supported by a similar observation
made recently in the analysis of sequences of ideal pulses 
\cite{cywin08,uhrig08,uhrig08err,pasin10a,wang13}.

\section{Scaling due to singular autocorrelation}
\label{sec:cusp}

In view of the general framework developed in Sect.\ \ref{sec:FN} we know that
we have to consider the average of $\vec{\mu}^2$ over the random configurations
$\eta(t)$, see Eq.\ \eqref{eq:FN_vs_mu2}. The Magnus expansion, see Subsect.\
\ref{subsec:AHT}, helps us to identify the leading contributions in 
an expansion in powers of $\tp$. For instance, it is obvious that $\mu_x^{(2)}$
in Eq.\ \eqref{eq:Magnus_mu2x} contributes to the average of $\Delta\F^2$
only in order $\tau_\mathrm{p}^4$ because $\mu_x^{(2)} \propto\tau_\mathrm{p}^2$ due to the double integral.
Thus the only candidate for contributions of lower order is
\begin{align}
\label{eq:average_muy_and_muz}
&\overline{\left(\mu_y^{(1)}\right)^2}+\overline{\left(\mu_z^{(1)}\right)^2}=
\nonumber\\
&\int_0^{\tau_\text p}\text d t_1 \int_0^{\tau_\text p}\text d t_2\ g(t_1-t_2)\cos\left[\psi(t_1)-\psi(t_2)\right]
\end{align}
which follows from the definition \eqref{eq:Magnuns_mu1} and 
the appropriate trigonometric relation. 

To understand the
behavior of the expression on the right hand side of \eqref{eq:average_muy_and_muz}
we use an expansion of the autocorrelation around $t=0$
\begin{equation}
\label{eq:Taylor_g_exp}
g(t)=g_0^2-g_1^2\gamma|t|+{\cal O}\left(t^2\right).
\end{equation} 
Inserting the leading order of this expansion in the integral of
Eq. (\ref{eq:average_muy_and_muz})  yields
\begin{subequations}
\begin{align}
\label{eq:I1_1}
I_1&:=g_0^2 \int_0^{\tau_\text p}\text d t_1 \int_0^{\tau_\text p}\text
d t_2\ \cos\left[\psi(t_1)-\psi(t_2)\right] \\
\label{eq:I1_2}
& =g_0^2\left[\int_0^{\tau_\text p}\text dt\cos\psi(t)\right]^2+g_0^2\left[\int_0^{\tau_\text p}\text dt\sin\psi(t)\right]^2
\end{align} 
\end{subequations}
which implies $\Delta\F^2=\frac{4}{3}(I_1+ \text{higher}\ \text{orders})$.
Clearly, this contribution vanishes if the condition \eqref{eq:first_cond} is fulfilled.
This means that this contribution vanishes for any first-order pulse. This is
perfectly consistent since $I_1$ would scale $\propto\tau_\mathrm{p}^2$ if it is non-vanishing so that
$\Delta\F\propto\tp$ would ensue.

Next, we consider the linear order in the expansion 
Eq.\ (\ref{eq:average_muy_and_muz}) yielding
\begin{align}
\label{eq:I1.5}
&I_{3/2}=
\\ \nonumber
&-g_1^2\gamma \int_0^{\tau_\text p}\text d t_1 \int_0^{\tau_\text p}\text
d t_2 |t_1-t_2|\cos\left[\psi(t_1)-\psi(t_2)\right]. 
\end{align} 
Clearly, this term scales like $I_{3/2}\propto \tau_\mathrm{p}^3\propto 1/v^3$ so that it 
readily explains the occurrence of $\Delta\F^2 \propto 1/v^3$
implying  $\Delta\F \propto 1/v^{3/2}$ naturally.
Therefore, the origin of the $3/2$ behaviour of the Frobenius norm has been identified.
A quantitative check of the prefactor in the scaling
is carried out for the SCORPSE and the CORPSE pulse in \ref{app:B}.
The quantitative agreement found supports our conclusion.

We underline that the contribution $I_{3/2}$ does not occur if
the autorcorrelation $g(t)$ is assumed to be analytical. Since $g(t)=g(-t)$
holds by definition, the analyticity prevents odd powers of $t$ to occur.
Then, no finite $g_1$ term occurs in the expansion \eqref{eq:Taylor_g_exp}
and thus no $I_{3/2}$ can arise. The next-leading contribution would
be a term $I_2\propto \tau_\mathrm{p}^4$.

We point out that the occurrence of $I_{3/2}$ due to the singularity of
the autorcorrelation at $t=0$ does not rely on the particular angle
of rotation studied, here $\pi$-pulses. Thus we expect that qualitatively
the same feature occurs for any control pulse.

In view of the above results one has to wonder whether we tried 
inappropriate pulses and whether it is not possible to 
make $I_{3/2}$ vanish by appropriate design of the pulse.
Indeed, we tried to find pulses also fulfilling $I_{3/2}=0$ besides the first-order conditions
\eqref{eq:first_cond}. But we failed to find such solutions. 
Instead we realized that is is rigorously impossible that such
pulses exist. We prove in \ref{app:C}
that the two standard first-order conditions \eqref{eq:first_cond} implying
$I_1=0$  and the additional condition $I_{3/2}=0$  cannot 
be fulfilled simultaneously. Note that this naturally explains
why all the pulses whose performance is simulated in Fig.\ \ref{fig:Frob_exp}
behave qualitatively the same, i.e., displaying the same exponent, except
for the non-refocusing rectangular pulse.

This is a strict no-go statement relevant for refocusing pulses
in presence of classical noise 
with a autocorrelation function with a cusp at $t=0$. 
As a consequence, pulses designed for 
this kind of
noise should be optimized in such a way that $I_{3/2}$ is minimized under
the constraint that the standard integrals 
\eqref{eq:first_cond} vanish. 
Then, the detrimental influence of the half-integer contribution $\Delta_\text
F\propto 1/v^{3/2}$ is reduced to its theoretical minimum.

We expect that this finding extends to noises with weaker singularities.
For instance, a singularity $\ln(g(t)) \propto |t|^3$ implies high-frequency
tails $\propto 1/\omega^4$. Note that recent spin noise measurements 
indeed revealed tails with exponents $\propto 1/\omega^{3.6}$ \cite{alvar11b} for
the noise experienced by the $^{13}C$ spins in adamantane.
We presume that such noise induces non-vanishing terms $I_{5/2}$ which
imply scaling exponents of $5/2$.

We point out that the fundamental limit for 
improving pulses applies also to the case where $T_1$ and $T_2$
do not differ substantially, e.g., in liquid NMR. 
This is so because an \textit{additional} coupling to the noise
clearly complicates the task to construct a refocusing pulse.
Indeed, additional constraints have to be fulfilled by
the pulses in linear and any higher order, see Ref.\ 
\cite{stihl12b}. But because the reduced set of
equations for the pure dephasing case with $T_1=\infty$
does not have a solution as proven here, no 
extended set will have a solution either.

Nevertheless, further studies of more specific models
are called for, for instance, it has to be elucidated how
the widespread case of spectral densities between Lorentzian and Gaussian
shape \cite{vleck48,beckm88} influence the possibility to construct 
efficient refocusing pulses.

\section{Conclusions}
\label{sec:conclusions}

In this paper, we have studied a spin $S=1/2$ subject to random dephasing
due to classical noise of Gaussian statistics. We numerically simulated 
and analytically analysed the performance of various spin-flip pulses.
To this end, we have introduced a mathematically clearly defined measure
for the quality of the pulses, namely the Frobenius norm $\Delta\F$ of the difference
between the ideal outcome of an ideal pulse and the real outcome of the 
real pulse of finite duration. The Frobenius norm is closely related to the
fidelity.

The aim was to show how pulses shaped to refocus the detrimental dephasing
perform and that this can be seen in the scaling of $\Delta\F$
with the pulse duration $\tp$. A standard rectangular pulse shows
 $\Delta\F\propto \tp$, a first-order pulse $\Delta\F\propto \tau_\mathrm{p}^2$, 
a second-order pulse $\Delta\F\propto \tau_\mathrm{p}^3$,  and so on. This could indeed be
verified in the simulations assuming noise with analytical autocorrelation $g(t)$.
But the simulation of noise with singular autocorrelation displayed unexpected
behavior. Such autocorrelations are generic for Ornstein-Uhlenbeck processes, that means for systems with spectral densities of Lorentzian shape.
We simulated a process which is governed by
$g(t)\propto \exp(-\gamma|t|)$, i.e., with a cusp at $t=0$.
 The pulse shaping does not improve the scaling beyond first order.
All refocusing pulses, i.e., shaped pulses beyond the rectangular pulses,
display anomalous scaling $\Delta\F\propto \tau_\mathrm{p}^{3/2}$.

We have clarified analytically that the anomalous scaling stems from the
cusp at $t=0$. It could be rigorously proven that no pulse design
can avoid the anomalous scaling.

We stress that a cusp at $t=0$ is not just a mathematical peculiarity
because the cusp is generated by the coupling to high-frequency fluctuations.
Thus, the deeper physical reason for the impossibility to improve the
pulse performance by shaping is that infinitely fast fluctuations contribute 
to the noise. They are manifest in the spectral density of cusp-like 
autocorrelations displaying high-frequency tails $\propto 1/\omega^2$
which do not converge quickly, i.e., exponentially, to zero.
Similar findings have been made before in the analysis in the design 
of sequences of ideal pulses \cite{cywin08,uhrig08,uhrig08err,pasin10a,wang13}.
The sequences could not be improved beyond a certain scaling if
the power spectrum of the noise displayed soft high-frequency cutoffs
following power laws.

We stress that 
processes with cusps in the autocorrelation function are common in nature. 
For instance, the exponential autocorrelation belongs to Ornstein-Uhlenbeck noise 
which is observed wherever a slow decay into a continuum of states without 
high-energy cutoff occurs. Note that on the frequency scale of microwaves 
already translational and rotational motion takes place fast, i.e.,
on practically infinitesimally short time scales. Thus the coupling to such
degrees of freedom induces homogeneous broadening with Lorentzian spectral density
\cite{abrag78}.

Summarizing, the results reported here imply an important general
message which is
largely independent of the details of the system:
There are fundamental limits to pulse refocusing if the noise
comprises significant high-frequency components. 
We investigated the simplest model where this
phenomenon occurs so that the effect is not obscured
by complexity. Thus we restricted ourselves to $T_1=\infty$ 
considering only spin-spin relaxation. But our conclusions apply 
also to the case where $T_1$ and $T_2$ are comparable
because the inclusion of additional couplings to random noise
\emph{increases} the number of constraints to be fulfilled by refocusing
pulses. Hence the impossibility of higher order refocusing pulses
for infinite $T_1$ implies the impossibility of such
pulses for finite $T_1$.

These observations may be seen as bad
news, but there is also some good news for the analysis of noise. 
If anomalous scaling is found in the performance of refocusing pulses
one can deduce that the noise is characterized by singular autorcorrelations.
These observations should be of use in a large variety of NMR experiments.

\section{Acknowledgements}
We are thankful to Roland B\"ohmer and Dieter Suter for helpful discussions
on relaxation mechanisms in NMR.
We acknowledge financial support by the Studienstiftung des deutschen Volkes (DS)
and by the DFG in project UH 90/9-1 (GSU).

\appendix 

\section{Piecewise constant pulses}
\label{app:A}

Here we present the details of the $\pi$-pulses simulated in Sect.\ \ref{sec:simulation}. 
All pulses are piecewise constant and rotate the spin around the $\sigma^x$-axis or any other
fixed axis in the $\sigma^x\sigma^y$-plane.  The investigated pulses comprise 
the symmetric SCORPSE and the asymmetric CORPSE pulse~\citep{cummi00,cummi03} which are
both first-order pulses. Furthermore, a second-order pulse (CLASS2ND) derived for a classical bath ~\citep{stihl12b}
 is studied. In addition, we discuss the second-order symmetric SYM2ND and the asymmetric ASYM2ND pulses 
derived in Ref.~\citep{pasin09a} where the bath was treated on the quantum level.

\begin{figure}[tb]
 \centering
  \includegraphics[width=0.7\columnwidth,angle=-90]{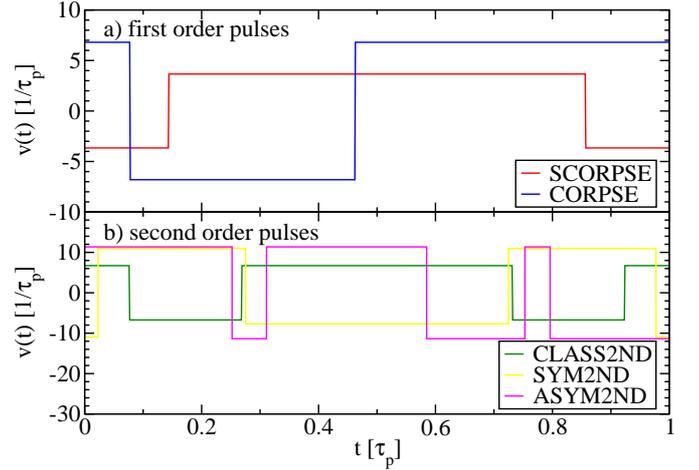}
   \caption{Pulse amplitude $v(t)$ versus time $t$. The first-order $\pi$-pulses are the 
symmetric SCORPSE and asymmetric CORPSE pulse~\citep{cummi00,cummi03}. 
In second order, a $\pi$-pulse (CLASS2ND) derived for a classical bath~\citep{stihl12b} 
and the quantum mechanical symmetric SYM2ND and asymmetric ASYM2ND $\pi$-pulses~\citep{pasin09a} are presented.
}
\label{fig:app_pulses}
\end{figure}

The time dependence of the pulse amplitudes $v(t)$ is plotted in Fig.~\ref{fig:app_pulses}, 
while the corresponding switching instances $\tau_i$, at which the amplitudes
jump, and their values $v_{\tau_i}$ are listed  in Tab.~\ref{tab:app_pulses}. 

\renewcommand{\arraystretch}{1.25}
\begin{table}[tb]
  \centering
 \begin{tabular}{cc}
$v_{\tau_i}\ [1/\tp]$ & $\tau_i\ [\tp]$ \\ \hline\hline
  \multicolumn{2}{c}{SCORPSE} \\ \hline
  $\pm\frac{7\pi}{6}$ & $\frac{1}{7}$ \\
    & $\frac{6}{7}$ \\ \hline
    \multicolumn{2}{c}{CORPSE} \\ \hline
  $\pm\frac{13\pi}{6}$ & $\frac{1}{13}$ \\
    & $\frac{6}{13}$ \\ \hline
    \multicolumn{2}{c}{CLASS2ND} \\ \hline
  $\pm 6.72572865242397$ & $0.07623077665509$ \\
    & $0.26784318744464$ \\ 
     & $0.73215681255536$ \\
      & $0.92376922334491$ \\ \hline
      \multicolumn{2}{c}{SYM2ND} \\ \hline
  $\pm 10.95012043866828575$ & $0.0228054551625108$ \\
   $-7.69537638364247465$  & $0.2752692173069500$ \\ 
     & $0.7247307826930500$ \\
      & $0.9771945448374892$ \\ \hline
      \multicolumn{2}{c}{ASYM2ND} \\ \hline
  $\pm 11.36443379447147705$ & $0.2520112376736856$ \\
    & $0.3108959015038718$ \\ 
     & $0.5847810746672190$ \\
      & $0.7528254671237393 $ \\ 
      & $0.7960392449336322 $ \\ 
       \end{tabular}
  \caption{Amplitudes $v_{\tau_i}$ and switching instances $\tau_i$ of the first- and second-order $\pi$-pulses 
  used in the numerical simulations. The resulting amplitudes $v(t)$ are plotted in Fig.~\ref{fig:app_pulses}.}
  \label{tab:app_pulses}
\end{table}
\renewcommand{\arraystretch}{1}

The parameters of the second-order pulses in Tab.~\ref{tab:app_pulses} are given in the same 
high precision as used in our numerical simulation. This allows us to illustrate the 
power law scaling down to very small values of $1/v$ and of $\Delta\F$. If the accuracy in
a simulation or in an experimental realization is lower the power laws are cut off 
on the scale of the accuracy, see plateaus in Fig.\ \ref{fig:Frob_gaussian}
for illustration. This means that plateaus occur at values $\Delta\F$
of the order of the accuracy. For instance, if the accuracy is say $10^{-2}$,
$\Delta\F$ will not fall below about $10^{-2}$, if the accuracy is  $10^{-3}$,
$\Delta\F$ will form a plateau at about $10^{-3}$.

\section{Quantitative check of the anomalous scaling}
\label{app:B}

In Sect.~\ref{sec:cusp}, the analytic expression \eqref{eq:I1.5} for $I_{3/2}$
has been derived which explains the anomalous scaling $\Delta\F \propto \tau_\mathrm{p}^{3/2} \propto 1/v^{3/2}$.
To be sure that this is the complete explanation we here check $\Delta\F^2 = (4/3)I_{3/2}+{\cal O}(\tau_\mathrm{p}^4)$
quantitatively for some pulses. 
The following explicit results apply to 
the first-order CORPSE and SCORPSE pulse, but the extension to other 
pulses is straightforward.

\begin{figure}[tb]
 \centering
\includegraphics[width=0.7\columnwidth,angle=-90]{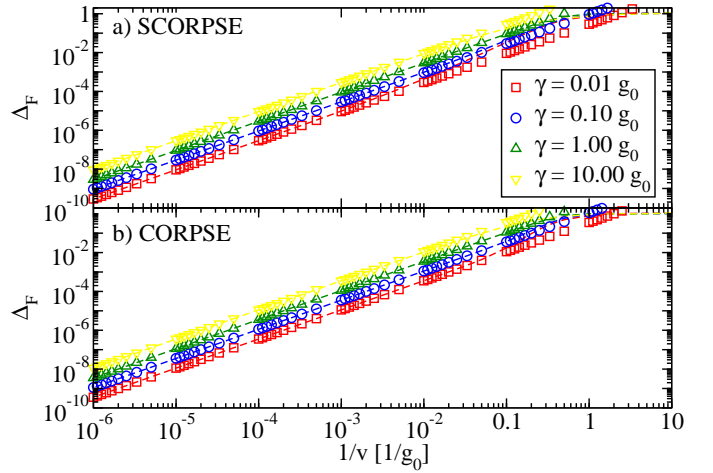}
  \caption{Frobenius norm $\Delta\F$ versus the inverse pulse amplitude $1/v$ for the autocorrelation $g(t)=g_0^2\e^{-\gamma |t|}$. The analytic results (dashed lines) quantitatively explain the unexpected
  $\Delta\F \propto 1/v^{3/2}$ behavior including the prefactor. The numerical results are depicted
  by the symbols. Note that there is no fit involved.}
  \label{fig:06_AHT}
\end{figure}
In order to check $\Delta\F^2 = (4/3)I_{3/2}$,
we have to carry out the integrals in Eq.~\eqref{eq:I1.5}.
The evaluation for the individual pulses is straightforward,  but rather lengthy. We only give a 
short description of the procedure and present the results.
One has to evaluate the time-dependent angle $\psi(t)=2\int_0^t\dt'\,v(t')$ which 
for the first-order CORPSE and SCORPSE pulses reads
\begin{subequations}
\begin{align}
 \psi_{\mathrm{CORPSE}}\left(t\right)&=\begin{cases} \frac{13\pi}{3\tau_\mathrm{p}}\cdot t & 0 \le t \le \frac{\tau_\mathrm{p}}{13} 
 \\
      \frac{2\pi}{3}-\frac{13\pi}{3\tau_\mathrm{p}}\cdot t & \frac{\tau_\mathrm{p}}{13}< t \le \frac{6\tau_\mathrm{p}}{13} 
      \\
      -\frac{10\pi}{3}+\frac{13\pi}{3\tau_\mathrm{p}}\cdot t & \frac{6\tau_\mathrm{p}}{13}< t \le \tau_\mathrm{p} 
      \\
     \end{cases}
\end{align}
\end{subequations}
and
\begin{subequations}
\begin{align}
 \psi_{\mathrm{SCORPSE}}\left(t\right)&=\begin{cases} -\frac{7\pi}{3\tau_\mathrm{p}}\cdot t & 0 \le t \le \frac{\tau_\mathrm{p}}{7} 
 \\
      -\frac{2\pi}{3}+\frac{7\pi}{3\tau_\mathrm{p}}\cdot t & \frac{\tau_\mathrm{p}}{7}< t \le \frac{6\tau_\mathrm{p}}{7} 
      \\
      \frac{10\pi}{3}-\frac{7\pi}{3\tau_\mathrm{p}}\cdot t & \frac{6\tau_\mathrm{p}}{7}< t \le \tau_\mathrm{p} 
      \\
     \end{cases},
\end{align}
\end{subequations}
respectively. Inserting these expressions in Eq.~\eqref{eq:I1.5} allows us to carry out
 the two-dimensional integration analytically. Finally, this yields the power laws 
\begin{subequations}
\label{eq:06_power_1st}
 \begin{align}
  \left.\Delta\F^2\right|_\mathrm{CORPSE\phantom{S}}&=4\pi g_0^2\gamma\left(\frac{1}{v}\right)^{3}\\
  \left.\Delta\F^2\right|_\mathrm{SCORPSE}&=\frac{8\pi}{3} g_0^2\gamma\left(\frac{1}{v}\right)^{3}
 \end{align}
\end{subequations}
for the leading order of the Frobenius norm $\Delta\F$.
In Fig.~\ref{fig:06_AHT}, these power laws (dashed lines) are compared with the numerical
results  from simulation. For both pulses and all values of $\gamma$, the agreement is perfect.
 Hence, the anomalous scaling $\Delta\F~\propto 1/v^{3/2}$ can indeed be attributed to the contribution 
$I_{3/2}$ given in  Eq.~\eqref{eq:I1.5}.

\section{Proof of the no-go theorem}
\label{app:C}

Here we prove that $I_1$ in Eq.~\eqref{eq:I1_2}  and $I_{3/2}$ in Eq.~\eqref{eq:I1.5}
cannot be made vanish simultaneously. To this end, we rewrite 
\begin{subequations}
\begin{align}
  I_{3/2} &= -a I_{3/2}^a  - a  I_{3/2}^b 
 \\
  I_{3/2}^a  &= \int\limits_0^\tp\dt_1\int\limits_0^\tp\dt_2\,\cos\psi_1\left|t_1-t_2\right| \cos\psi_2 
  \\
  I_{3/2}^b &= \int\limits_0^\tp\dt_1\int\limits_0^\tp\dt_2\,\sin\psi_1\left|t_1-t_2\right| \sin\psi_2,
\end{align}
\end{subequations}
where $a=g_1^2\gamma$ and we have abbreviated $\psi_i:=\psi(t_i)$ for brevity.
In the Hilbert space $\mathscr{H}$ of real integrable functions in the time interval $[0,\tp]$
 both integrals can be interpreted as expectation values 
\begin{subequations}
 \begin{align}
   I_{3/2}^a  &= \braket{\cos(\psi)|A|\cos(\psi)} \\
   I_{3/2}^b &= \braket{\sin(\psi)|A|\sin(\psi)}
 \end{align}
\end{subequations}
of the linear operator $A$ with the matrix elements $\left|t_1-t_2\right|$.
The corresponding mapping of the function $\varphi(t)\in\mathscr{H}$ to the function $\psi(t)$ 
reads
\begin{align}
 A:\quad &\varphi\left(t\right)\longmapsto \psi\left(t\right)=\int\limits^\tp_0 \dt'\, A(t,t') \varphi(t').
\end{align}

The idea is to write the operator $A$ in a form containing $B^\dagger B$
because this implies that $I_{3/2}$ is non-negative.
An educated guess for an absolute value is the sign function because it is
the derivative of the absolute value function. Thus we consider
\begin{align}
  B:\quad &\varphi\left(t\right)\longmapsto \psi\left(t\right)=\int\limits^\tp_0 \dt'\, \sgn(t-t') \varphi(t'). 
\end{align}
Obviously, this operator is antihermitian, i.e., $B^\dagger=-B$.
We compute the square $B^2$
\begin{align}
\nonumber
  B^2:\quad &\varphi\left(t\right)\longmapsto \chi\left(t\right) =\int\limits_0^\tp \dt_1\int\limits_0^\tp\dt_2\, \sgn(t-t_1)  
  \\
 &\quad \times\sgn(t_1-t_2) \varphi(t_2). 
\end{align}
The integration with respect to $t_1$ can be carried out analytically
leading to
\begin{align}
  B^2:\quad &\varphi\left(t\right)\longmapsto \chi\left(t\right)=\int\limits^\tp_0 \dt_2\, \left(2\left|t-t_2\right|-1\right) \varphi(t_2). 
\end{align}
Indeed, this establishes a relation between $A$ and $B^2=-B^\dagger B$ which reads
\begin{align}
 A&=\frac{1}{2}\left(C-B^\dagger B\right)
 \label{eq:app_A}
\end{align}
where we used $B=-B^\dagger$ and introduced the operator 
\begin{align}
   C:\quad &\varphi\left(t\right)\longmapsto \psi\left(t\right)=\int\limits^\tp_0 \dt'\, \varphi(t').
\end{align}

Next, we observe that a pulse with $\psi(t)$ which fulfills the first-order condition
\eqref{eq:first_cond} automatically obeys 
\begin{subequations}
\begin{eqnarray}
0 &=& C|\sin(\psi)\rangle
\\
0 &=& C|\cos(\psi)\rangle .
\end{eqnarray}
\end{subequations}
Thus, in the subspace of first-order pulses the relations
\begin{subequations}
\begin{eqnarray}
 I_{3/2}^a &=& -\frac{1}{2} \langle B\cos(\psi)|B \cos(\psi)\rangle \le 0
\\
 I_{3/2}^b &=& -\frac{1}{2} \langle B\sin(\psi)|B \sin(\psi)\rangle \le 0 
 \\
 I_{3/2} &=& \frac{a}{2}\left( \langle B\cos(\psi)|B \cos(\psi)\rangle \right. \nonumber
 \\
  && \quad + \left. \langle B\sin(\psi)|B \sin(\psi)\rangle \right)
 \ge 0 \qquad 
\end{eqnarray}
\end{subequations}
hold demonstrating the non-negativity of the anomalous contribution.
Hence $I_{3/2}=0$ implies that $0=B|\cos(\psi)\rangle = B|\sin(\psi)\rangle$.
Functions $\varphi(t)=\cos(\psi(t))$ or $\varphi(t)=\sin(\psi(t))$ which are annihilated by the mapping $B$ fulfill
\begin{equation}
0=  \int\limits^\tp_0 \dt'\, \sgn(t-t') \varphi(t')  
\label{eq:app_exp_A0}
\end{equation}
for all values of $t$. Taking the derivative with respect to $t$ on both sides
implies
\begin{equation}
 0 =2\int\limits^\tp_0 \dt'\, \delta(t-t') \varphi(t') =2\varphi(t)
\end{equation}
so that $\varphi(t)$ has to vanish for all $t$. No such solution exists for $\psi(\tp)-\psi(0)=\pi$.
We conclude that it is impossible to make $I_1$ and $I_{3/2}$ zero simultaneously.

As a consequence, pulses designed for Ornstein-Uhlenbeck noise 
can only be optimized in such a way that $I_1=0$, see Eq.\ \eqref{eq:first_cond},
and that $I_{3/2}$ is minimized. This is the best one can aim for.


\end{document}